\documentclass[prl,showpacs,onecolumn]{revtex4}
\usepackage{dcolumn}
\usepackage{bm}
\usepackage{amssymb}
\usepackage{amsmath}
\usepackage{amsfonts}
\usepackage[]{graphicx}
\begin{document}
\bibliographystyle{prsty}
\title{ Can quantum mechanics be considered as statistical? }
\author{Aur\'{e}lien Drezet}
\affiliation{Institut N\'eel UPR 2940, CNRS-University Joseph Fourier, 25 rue des Martyrs, 38000 Grenoble, France}

\date{\today}

\begin{abstract}
This is a short manuscript  which was  initially submitted to Nature
Physics as a comment to the PBR (Pusey, M.~F., Barrett, J., Rudolph,
T) paper just after its publication in 2012. The comment was not
accepted. I however think that the argumentation is correct: one is
free to judge!
\end{abstract}

\maketitle
\textbf{To the Editor}- Despite so many successes quantum mechanics is still nowadays steering intense interpretational debates. In this context Pusey, Barrett and Rudolph (PBR) presented recently a new 'no-go' theorem~\cite{PBR} whose aim is to restrict drastically the family of viable quantum interpretations. For this purpose PBR focussed on what Harrigan and Speckens~\cite{speckens} named `epistemic' and `ontic' interpretations~\cite{news,news2} and showed that only the second family can agree with quantum mechanics.\\
\indent Here, we analyze the PBR theorem and show that while mathematically true its correct physical interpretation doesn't support the conclusions of the authors.\\
\indent In the simplest version PBR considered two non orthogonal pure quantum states $|\Psi_1\rangle=|0\rangle$ and $|\Psi_2\rangle=[|0\rangle+|1\rangle]/\sqrt{2}$ belonging to a 2-dimensional Hilbert space $\mathbb{E}$ with basis vectors $\{|0\rangle,|1\rangle\}$. Using a specific measurement basis $|\xi_i\rangle$ ($i\in[1,2,3,4]$) in  $\mathbb{E}\otimes\mathbb{E}$  (see their equation 1 in \cite{PBR}) they deduced that $\langle\xi_1|\Psi_1\otimes\Psi_1\rangle=\langle\xi_2|\Psi_1\otimes\Psi_2\rangle=\langle\xi_3|\Psi_2\otimes\Psi_1\rangle=\langle\xi_4|\Psi_2\otimes\Psi_2\rangle=0$. In a second step they introduced hypothetical `Bell's like' hidden variables $\lambda$ and wrote implicitly the probability of occurrence in the form:
\begin{eqnarray}
|\langle\xi_i|\Psi_j\otimes\Psi_k \rangle|^2=\int\int P(\xi_i|\lambda,\lambda')\rho_j(\lambda)\rho_k(\lambda')d\lambda d\lambda'
\end{eqnarray}
where $i\in[1,2,3,4]$ and $j,k\in[1,2]$. In this PBR model there is a independence criterion at the preparation since we write $\rho_{j,k}(\lambda,\lambda')=\rho_j(\lambda)\rho_k(\lambda')$. In these equations we introduced the conditional `transition' probabilities $P(\xi_i|\lambda,\lambda')$
for the outcomes $\xi_i$ supposing the hidden state
$\lambda,\lambda'$ associated with the two independent Q-bits are given. Obviously, we have  $\sum_{i=1}^{i=4}P(\xi_i|\lambda,\lambda')=1$. It is then easy using all these definitions and conditions to  demonstrate that we must necessarily have $\rho_2(\lambda)\cdot\rho_1(\lambda)=0$ for every $\lambda$ i.e. that $\rho_1$ and $\rho_2$ have nonintersecting supports in the $\lambda$-space. This constitutes the PBR theorem for the particular case of independent prepared states $\Psi_1,\Psi_2$ defined before. PBR generalized theirs results for more arbitrary states using similar and astute procedures described in ref.~1.\\ The general PBR theorem states that the only way to include hidden variable in a
description of the quantum world is to suppose that for every pair
of  quantum states $\Psi_1$ and $\Psi_2$ the density of
probability must satisfy the condition of non intersecting support
in the $\lambda$-space:
\begin{eqnarray}
\rho(\lambda,\Psi_1)\rho(\lambda,\Psi_2)=0  & \forall \lambda.
\end{eqnarray}
If this theorem is true it would make hidden variables
completely redundant since it could be possible to
define a bijection or relation of equivalence between the $\lambda$
space and the Hilbert space: (loosely speaking we could in
principle make the correspondence $\lambda\Leftrightarrow\psi$).
Therefore it would be as if $\lambda$ is nothing but a new name
for $\Psi$ itself. This would justify the label `ontic' given to this kind of interpretations by opposition to `epistemic' interpretations ruled out by the PBR result.\\
However, this conclusion is wrong as it can be shown by examining carefully the assumptions necessary for the derivation of the theorem. Indeed, using the well known Bayes-Laplace formula for conditional probability we deduce that the most general Bell's hidden variable probability space should obey the following rule
\begin{eqnarray}
|\langle\xi_i|\Psi_j\otimes\Psi_k \rangle|^2=\int\int P(\xi_i|\Psi_j,\Psi_k,\lambda,\lambda')\rho_j(\lambda)\rho_k(\lambda')d\lambda d\lambda'
\end{eqnarray}
in which in contrast with equation 1 the transition probabilities $P(\xi_i|\Psi_j,\Psi_k,\lambda,\lambda')$ now depend explicitly on the considered quantum states $\Psi_j,\Psi_k$. Relaxing this PBR assumption has a direct effect since we loose the ingredient necessary for the demonstration of equation 2 (more precisely we are not anymore allowed to compare the product states $|\Psi_j\otimes\Psi_k \rangle$ as it was done by PBR). In other words the PBR theorem collapses.\\
\indent Physically speaking our conclusion is sound since many hidden variable models and in particular the one proposed by de Broglie and Bohm belong to the class where the transition probabilities and the trajectories depend contextually  on the quantum states $\Psi$.\\
\indent To conclude, contrary to the PBR claim the theorem they
proposed is actually limited to a very narrow class of quantum
interpretations. It fits well with the XIX$^{th}$ like hidden
variable models using Liouville and Boltzmann approaches (i.e.
models where the transition probabilities are independent of $\Psi$)
but it is not in agreement with neo-classical theories such as the
one proposed by de Broglie and Bohm in which the wavefunction is at
the same time an epistemic and ontic ingredient of the dynamics.
Therefore epistemic and ontic states can not be separated in quantum
mechanics.\\
\indent \underline{\emph{additional remark not included in the
original text:}} The interested reader can also consider my two
papers on this subject in  arXiv:1203.2475 and arXiv:1209.2565
[Progress in Physics, vol 4 (October 2012)].

\end{document}